\documentclass[aps,prb,superscriptaddress,showpacs,a4paper,longbibliography]{revtex4-2}
\usepackage[utf8]{inputenc}
\usepackage{graphicx}
\usepackage{color}
\usepackage{fullpage}
\usepackage{amsfonts}
\usepackage{amsmath}
\usepackage{color}
\usepackage{mathtools}
\usepackage{booktabs}
\usepackage{enumitem}
\usepackage{longtable}
\usepackage{multirow}
\usepackage{rotating}
\usepackage{siunitx}
\usepackage{comment}
\usepackage{hyperref}
\usepackage[table]{xcolor}
\usepackage{makecell}
\newcolumntype{P}[1]{>{\centering\arraybackslash}p{#1\textwidth}}
\hypersetup{breaklinks=true,colorlinks=true}
\usepackage[table]{xcolor}
\usepackage{array}

\definecolor{Acolor}{RGB}{225, 200, 255} % rose/violet
\definecolor{Lcolor}{RGB}{180, 235, 180} % vert

\newcommand{\Agroup}{\rowcolor{Acolor}}
\newcommand{\Lgroup}{\rowcolor{Lcolor}}
\newcommand{\Nocolor}{} % mixte A+L+H

\usepackage{ulem}
\newcommand{\expred}[1]{\textbf{\textcolor{red!70!black}{#1}}}
\newcommand{\expblue}[1]{\textbf{\textcolor{blue!70!black}{#1}}}
\newcommand{\chk}{\textcolor{red}{\checkmark}}
\begin{document}

\title{Library of peptides: Analysis of a complex reactive mixture with the use of statistical physics and inference }

\author{Laura Messio}
\email{laura.messio@sorbonne-universite.fr}
\affiliation{Sorbonne Universit\'e, CNRS, Laboratoire de Physique Th\'eorique de la Mati\`ere Condens\'ee, LPTMC, F-75005 Paris, France}
\author{Taleen Peker}
\affiliation{Sorbonne Universit\'e, École Normale Supérieure, PSL University, CNRS, Laboratoire des biomolécules, LBM, F-75005 Paris, France}
\author{Roba Moumné}
\email{roba.moumne@sorbonne-universite.fr}
\affiliation{Sorbonne Universit\'e, École Normale Supérieure, PSL University, CNRS, Laboratoire des biomolécules, LBM, F-75005 Paris, France}
\author{Hélène Berthoumieux}
\email{helene.berthoumieux@espci.fr}
\affiliation{UMR CNRS Gulliver 7083, ESPCI Paris, PSL Research University, 75005 Paris, France}
\date{\today}
\begin{abstract}
Self-assembled mixtures enable the exploration of vast chemical diversity, yet characterizing the hundreds of coexisting species they contain remains a major challenge. Standard analytical techniques such as chromatography and spectroscopy produce overlapping signals that cannot be resolved by conventional deconvolution methods, which neglect the concentration correlations imposed by the reactive network. Here, we introduce a framework that bridges statistical physics and analytical chemistry to deconvolute complex self-assembled mixtures from partial spectral data. Inspired by Potts models of protein folding, we express the chemical potential of each compound as a function of a reduced set of parameters—intrinsic building-block energies and pairwise coupling terms between neighboring building blocks in the sequence. Coupling this model with the law of mass action yields equilibrium concentrations across the entire reactive network. The model parameters can be infered by fitting predicted spectroscopic signals to experimentally measured partial data. We demonstrate this approach using dynamic combinatorial libraries (DCLs) of peptides as a model system. Applied to a DCL of 81 peptides, our method resolves the full mixture composition from just 13 overlapping peaks, revealing that lateral couplings dominate the assembly energetics. Beyond DCLs, this strategy offers a general route for analyzing self-assembled systems from partial data and extends naturally to target-driven identification of high-affinity binders.
\end{abstract}
\maketitle
%\tableofcontents

\section{Introduction}
\begin{figure}
 	\begin{center}
 		\includegraphics[scale=0.7]{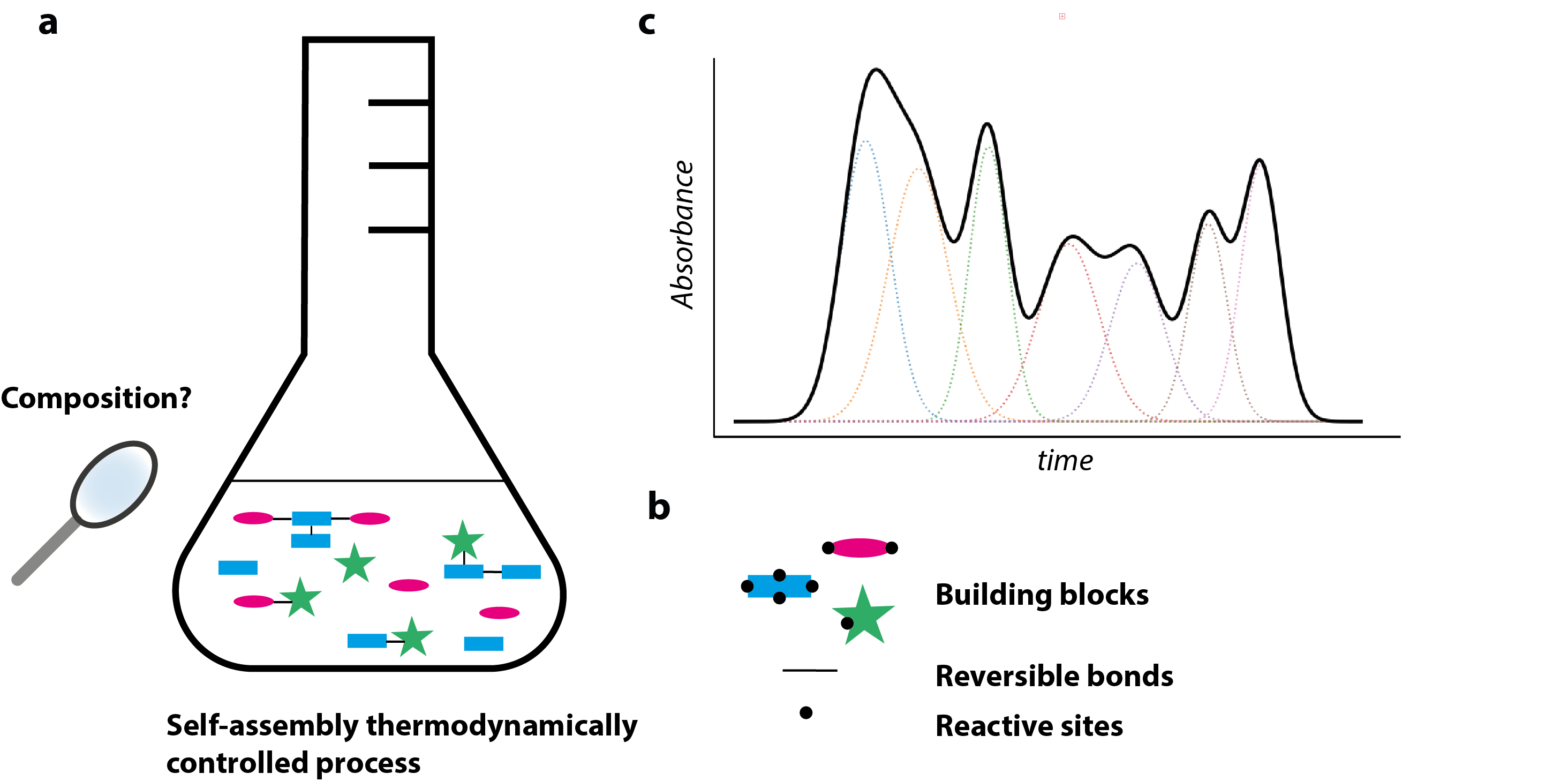}
 		\caption{a. Sketch of a self-assembly thermodynamically controlled mixture. b. Building blocks and reactive sites.  c. Associated experimental spectrum (solid black line) composed of 9 overlapping peaks (dashed color lines).}
 	\label{fig:sketch}
 	\end{center}
 \end{figure}

Identifying new active compounds—from biomolecules to 'smart' materials—is one of the major tasks of chemists. A powerful approach relies on dynamic self-assembly processes which allow to explore molecular diversity.\cite{lehn1997,steed} As illustrated in Fig.~\ref{fig:sketch}, this now-standard approach consists in mixing building blocks that can assemble via reversible covalent or non-covalent bonds, leading to the formation of hundreds of compounds. This allows exploration of a vast chemical phase space. However, this strength can become a weakness when it comes to characterizing the resulting mixture, a limitation that has led researchers to abandon certain approaches in the past. Attention has therefore shifted toward self-assembly at larger scales.\cite{doi:10.1126/science.1070821,doi:10.1073/pnas.082065899} Yet self-assembly at the molecular level remains the scale at which life emerges, and the analysis of such mixtures must be targeted. How can the complex assemblies of reactive species obtained through this approach be characterized?

For combinatorial mixtures, standard techniques such as chromatography and spectroscopy often produce spectra with overlapping peaks (see Fig.~\ref{fig:sketch} c), making their interpretation difficult—or even impossible.\cite{misucara}
To overcome this difficulty, peak deconvolution can be performed, based on a linear approximation and \textit{ad hoc} expressions for an individual peak (typically assuming a Gaussian shape) and its relative weight, which are fitted to maximise agreement with the experimental spectra.\cite{brouwer1973deconvolution,lacey1986deconvolution}

These spectra can also be deconvoluted using experimental techniques that increase the dimension of analysis; one example is two-dimensional nuclear magnetic resonance (2D NMR)~\cite{giusepponne2008,Morris1997,Nilsson2010}, but the number of species that can be targeted remains small compared to the phase space. 
Unsupervised methods such as simple component analysis (PCA)~\cite{surface_raman,ML-Raman,Jolliffe2002} or machine learning~\cite{Misuraca2020} can be employed for deconvolution of complex spectra, but these require large training datasets or extensive spectral replication, which is not always feasible.
In dynamic combinatorial libraries (DCLs), where hundreds of species coexist\cite{Otto1997,Cruciani2017}, this limitation is particularly acute.

A key limitation of conventional deconvolution approaches is that they generally treat the contributions of individual species as independent fitting parameters. Yet, in many combinatorial systems, the possible species are not independent: their abundances are linked by the underlying network of interactions and by the physical constraints governing the system. Incorporating these relationships into the analysis could therefore provide additional information, making it possible to infer the composition of a complex mixture even when the experimental data alone are insufficient to resolve its individual components.\cite{10.1021/j100051a012,berthoumieux2009,zrelli2011,querard2017} 

 To address this, we propose a new approach specially constructed to deconvolute spectra of self-assembly libraries. It takes into account the combinatorial nature of the ensemble of chemical species and the concentration correlations which are imposed by the reactive network, that we assume to be known. 

Statistical physics provides efficient frameworks for exploring such large configuration spaces made of repetitive blocks with a reduced number of parameters. A beautiful example of this is the study of protein folding. The phase space of the 3D structure of proteins is gigantic. However, the energy landscape can be described with a reduced number of parameters using the Potts model.\cite{LEVY201755} This is a spin model with q discrete possible values; the case q=2 corresponds to the Ising model. The energy of each conformer is expressed as a function of the intrinsic energies of the amino acids and of coupling terms between neighbors, encoding the spatial proximity between amino acids. We extend this approach to express the energy of self-assembled compounds as a function of the energies of building blocks and coupling terms illustrating the spatial arrangements of the building blocks. This allows expressing the energy of all possible compounds in the mixture using a small number of parameters. Furthermore, the concentrations of species within a reactive network are correlated by the law of mass action. Here, we express the abundance of all compounds as a function of their relative energy, the reactive network topology, and the initial concentrations of building blocks. Finally, we can express the concentrations of all species as a function of a small number of parameters, which we adjust to reproduce partial experimental data from the spectra.

\begin{figure}
	\begin{center}
		\includegraphics[width=\textwidth]{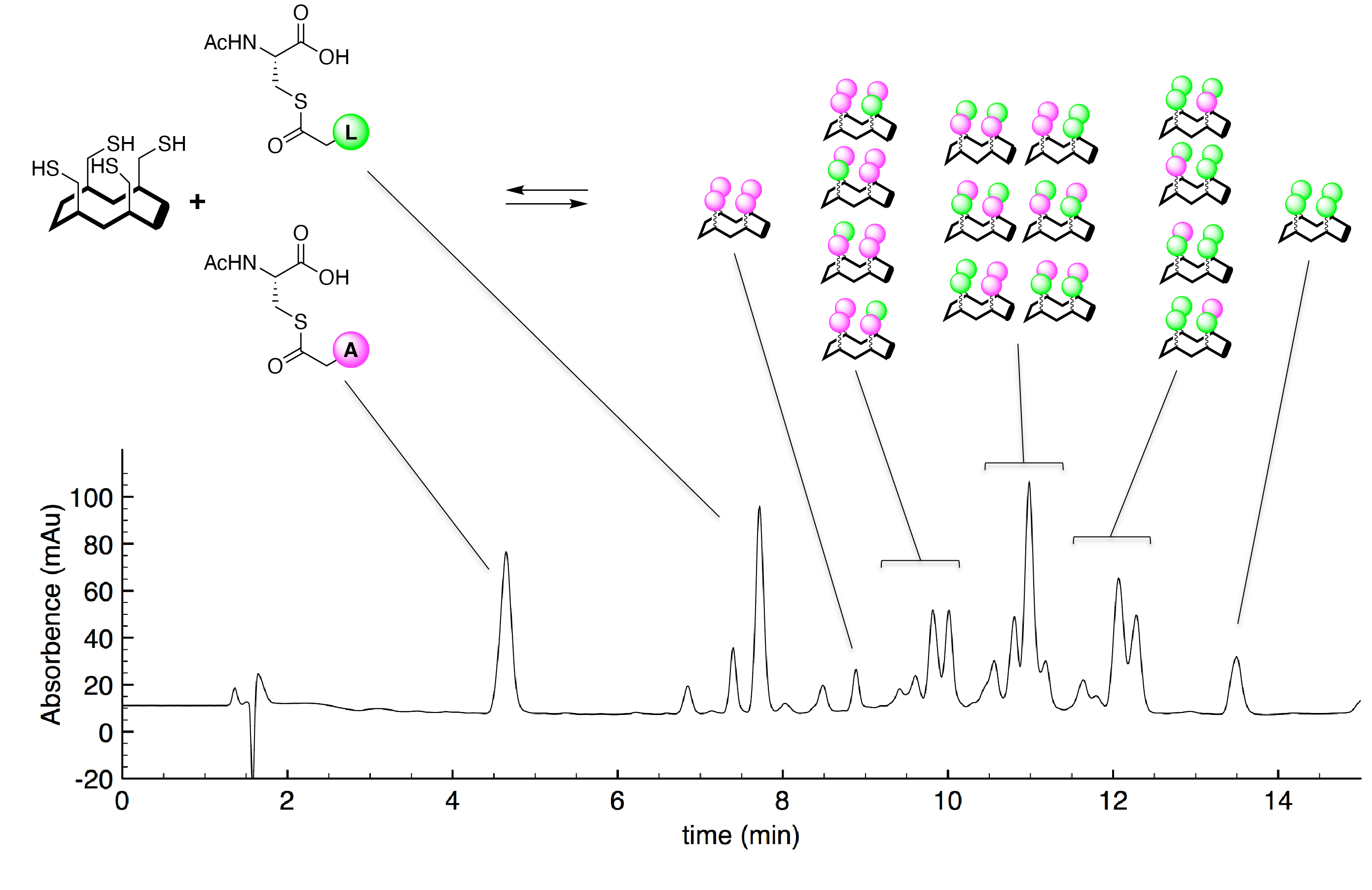}
		\caption{HPLC spectra for the experiments (25) detailed in Tab.~\ref{tab:experiments}. We can identify 5 peaks (or group of peaks ) in a mixture containing 81 species. 
			Therefore, the composition of the mixture is thus not completely characterised.
			\label{fig:spectra}
		}
	\end{center}
\end{figure}

We demonstrate our method using a DCL of peptides.\cite{DCL2019,dcl_rev,chem.202200454}  In this system, a scaffold with four reactive sites forms reversible bonds with a set of two building blocks (BBs)—in this case, the amino acids alanine and leucine introduced in excess. The library (DCL) contains 81 molecules and is at a stable thermodynamic equilibrium.
The Figure~\ref{fig:spectra} shows the  RP-HPLC (Reversed-Phase High-Performance Liquid
Chromatography) for the mixture at equilibrium, evidencing the sub-characterization of the system.

Our approach enables the determination of the full composition of this DCL, from this spectrum. 
Beyond the specific system studied here, our results establish a strategy for analyzing complex reactive mixtures by bridging statistical physics and analytical chemistry. Potential applications extend beyond DCLs for all self-assembled systems.
This work is organized as follows: First, we present the experiments we consider in Sec.\ref{sec:expData}. We then describe the deconvolution methods in detail (Sec.\ref{sec:method}). The validity of our approach is confirmed by dividing our experimental data into a fitting set and a validation set, which we then validate. Finally, we determine the composition of the DCL.

\section{Experimental data}
\label{sec:expData}

\subsection{Notations}
\begin{figure}
\begin{center}
  \includegraphics[width=\textwidth]{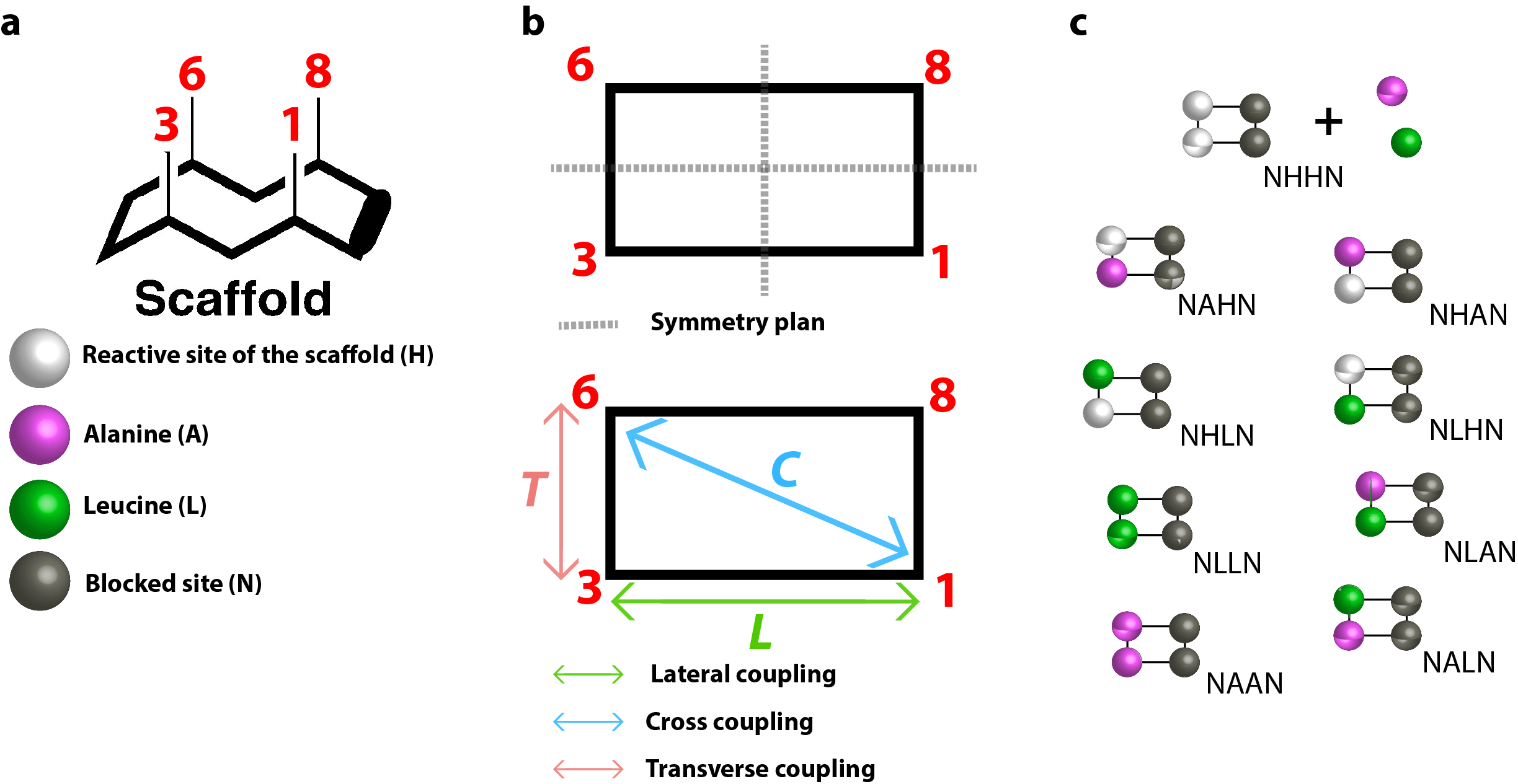}
  \caption{Dynamic combinary library of peptides. 
  (a) Sketch of the carbon backbone and possible reactive sites. 
  (b) Symmetry plane of the top face of the backbone and coupling terms considered in the model. 
  (c) Sketch of all the peptides formed in the reactive mixture for experiments (13-15) of \ref{tab:experiments} for the NHHN scaffold. 
  \label{fig:backbone}}
\end{center}
\end{figure}

%A theoretical model with a reduced set of parameters will be constructed. The concentrations of each peptide will then be calculated and compared to a set of {\it experiments}. The parameters will then be optimised to fit the experimental results. 

An \textit{experiment} consists of the initial composition data of a solution of scaffold and BB together with the data resulting from RP-HPLC for the mixture at equilibrium~\cite{chem.202200454} and giving partial information on the final composition.

The scaffold is a peptide presenting a reactive face composed of four  sites, labelled  $(1, 3, 6, 8)$, as shown in Fig.~\ref{fig:backbone}(a). 
These sites can be either reactive or blocked and are labelled H and N, respectively. In reality, H is not just a hydrogen atom, but a sulfur group~\cite{chem.202200454}, but we keep the simplified notation as it has no importance in the following.
The list $\mathcal B$ of blocked sites fully characterizes the scaffold. 
For example, $\mathcal B=\emptyset$ corresponds to HHHH, which has no blocked sites and four reactive sites, and $\mathcal B=\{1,8\}$ to NHHN, with two blocked and two reactive sites. 
The initial scaffold concentration is fixed at 1. 

The BB are amino acids. 
We denote $\mathcal A$ the set of amino acids in the solution, $n_{\rm BB}$ its cardinal, and $c^0_a$ the initial ratio of concentration with respect to the scaffold for $a\in\mathcal A$. 
Here, we consider solutions with alanine: $\mathcal A=\{A\}$, leucine $\mathcal A=\{L\}$ or a mix of these two amino acids $\mathcal A=\{A, L\}$. 
The BB are present in excess, with a concentration ratio that varies between 2 and 20. 

The solution reacts and evolves towards equilibrium. 
The peptides that can be found are obtained from $\mathcal B$ and $\mathcal A$ and constitute a set $\mathcal P^{\rm th}$.The Figure~\ref{fig:backbone}(c) presents  the set $\mathcal P^{\rm th}$ for $\mathcal B$=NHHN and $\mathcal A$=\{A,L\}.  
The relative composition of the mixture is monitored by RP-HPLC at equilibrium. 
Several peptides of $\mathcal P^{\rm th}$ can be associated to the same HPLC peak when the peaks overlap. 
 
Each peak thus corresponds to a group $p$ of peptides (with eventually a unique element). These groups form a set $\mathcal P^{\rm exp}$, of cardinal the number of peaks, smaller or equal to the cardinal of $\mathcal P^{\rm th}$  (see Tab.~\ref{tab:experiments}). 
%Then we cannot determine each peptide concentration, but only the sums of concentrations of elements of $\mathcal P^{\rm exp}$. 
The concentration $c_p$ associated with a peak (and to the group of peptides $p\in\mathcal P^{\rm exp}$) is proportional to its relative area. Note that the linearity of the detection method is assumed here. 
As we have fixed the initial scaffold concentration to 1, $c_p$ is a proportion, or the ratio of concentration of the sum of concentrations of a group $p$ of peptides with respect to the initial scaffold concentration. 

In summary, we mathematically denote an experiment by $\mathcal E = \lbrack \mathcal B, \{c^0_a, a\in\mathcal A\}, \{c_p, p\in\mathcal P^{\rm exp}\} \rbrack$,

\subsection{Experiment set}
We present 25 experiments in Tab.~\ref{tab:experiments} labeled by the first-column number. 
All of these experiments use the same peptide backbone, but use different scaffold: they present different combinations of protected and reactive sites indicated in the second column through the list of blocked sites $\mathcal B$. 
The third column gives the set $\mathcal A$ of amino acids and their initial concentrations. The fourth column gives the concentrations of peptides or groups of peptides identified from the HPLC-spectra analysis. For experiments 10 to 25, the proportion of each species is not accessible. Even for simple experiments involving 4 peptides such as (10-12), we only have access to partial information, for example the proportion of the two products (NAHN+NHAN). 
Note that the experimental protocol and the HPLC spectra can be found in Ref.~\cite{chem.202200454}.

As an example, we consider the reactions of experiment (13). 
The blocked sites are $\mathcal B=\{1,8\}$ and the reactive ones are $\{3,6\}$, thus the scaffold is NHHN. 
A reactive mixture contains initially the amino acids A and L, twice as concentrated as the backbone,   $\mathcal A =\{\rm A, \rm L\}$, $c_A^0=c_L^0=2$. 
In this case, the reactive mixture contains $3^2$=9 peptides, including the scaffold and all represented in Fig.~\ref{fig:backbone}(c).
HPLC spectra analysis (see Fig.~\ref{fig:spectra}) can distinguish  only 7 peaks for 9 scaffolds. One peak is a double peak corresponding to (NAHN, NHAN), which can not be distinguished and a second peak is associated to NLAN+NLHN. The concentrations of the peptides or groups of peptides associated to each peak are given in the last column of Tab.~\ref{tab:experiments}. 
\vspace{0.5cm}

\begin{longtable}{cccp{10cm}}
	\caption{Set of 25 experiments considered in this analysis. 
		The experiments labelled in blue compose a training set that we use for a validation of the method. The experiments in red compose a validation set, as explained in Sec.~\ref{sec:validation}.
		\textsuperscript{b} Scaffold in italic stands for all the possible combinations: {\it LHHH}=LHHH+HLHH+HHLH+HHHL. }
	\label{tab:experiments}\\
	\toprule
	Exp. & $\mathcal{B}$ & $c_a^0$ & Scaffold distribution (\%) \\
	\midrule
	\endfirsthead
	
	\toprule
	Exp. & $\mathcal{B}$ & $c_a^0$ & Scaffold distribution (\%) \\
	\midrule
	\endhead
	
	\bottomrule
	\endfoot
	
	\expred{1}  & $\{1,3,8\}$ & L (2)            & NNHN (32), NNLN (68) \\
	\expblue{2} & $\{1,3,8\}$ & L (6)            & NNHN (19), NNLN (81) \\
	\expblue{3} & $\{1,3,8\}$ & L (10)           & NNHN (13), NNLN (87) \\
	
	\expblue{4} & $\{1,3,8\}$ & A (2)            & NNHN (40), NNAN (60) \\
	\expred{5}  & $\{1,3,8\}$ & A (6)            & NNHN (22), NNAN (78) \\
	\expblue{6} & $\{1,3,8\}$ & A (10)           & NNHN (17), NNAN (83) \\
	
	\expblue{7} & $\{1,8\}$ & L (4)            & NHHN (13), NLHN (22), NHLN (16), NLLN (51) \\
	\expred{8}  & $\{1,8\}$ & L (6)            & NHHN (9), NLHN (20), NHLN (12), NLLN (57) \\
	\expblue{9} & $\{1,8\}$ & L (9)            & NHHN (7), NLHN (10), NHLN (7), NLLN (78) \\
	
	\expred{10} & $\{1,8\}$ & A (4)            & NHHN (19), NAHN+NHAN (47), NAAN (32) \\
	\expblue{11} & $\{1,8\}$ & A (6)           & NHHN (10), NAHN+NHAN (43), NAAN (47) \\
	
	\expblue{12} & $\{1,8\}$ & A (2), L (2)
	& NHHN (14), NAHN+NHAN (21), NHLN+NLHN+NLAN (32), NALN (8), NAAN (9), NLLN (16) \\
	
	\expred{13} & $\{1,8\}$ & A (3), L (3)
	& NHHN (9), NAHN+NHAN (18), NHLN+NLHN+NLAN (33), NALN (10), NAAN (9), NLLN (22) \\
	
	\expblue{14} & $\{1,8\}$ & A (4.5), L (4.5)
	& NHHN (4), NAHN+NHAN (13), NHLN+NLHN+NLAN (31), NALN (11), NAAN (10), NLLN (30) \\
	
	\expblue{15} & $\{1,6\}$ & L (4)            & NHNH (10), NLNH+NHNL (32), NLNL (58) \\
	\expred{16}  & $\{1,6\}$ & L (6)            & NHNH (2), NLNH+NHNL (20), NLNL (77) \\
	\expblue{17} & $\{1,6\}$ & L (9)            & NHNH (1), NLNH+NHNL (12), NLNL (86) \\
	
	\expblue{18} & $\{1,6\}$ & A (4)            & NHNH (16), NANH+NHNA (45), NANA (39) \\
	\expblue{19} & $\{1,6\}$ & A (6)            & NHNH (9), NANH+NHNA (40), NANA (51) \\
	\expblue{20} & $\{1,6\}$ & A (9)            & NHNH (3), NANH+NHNA (30), NANA (66) \\
	
	\expblue{21} & $\{1,6\}$ & A (2), L (2)
	& NHNH (12), NANH+NHNA (22), NANA (10), NLNH+NHNL+NLNA (28), NANL (9), NLNL (15) \\

	\expblue{22} & $\emptyset$ & L (40)
	& HHHH (0.5), {\it LHHH} (0.5), {\it LLHH} (3), {\it LLLH} (41), LLLL (55) \\
	
	\expred{23} & $\emptyset$ & L (10)
	& HHHH (2), {\it LHHH} (6), {\it LLHH} (22), {\it LLLH} (47), LLLL (22) \\
	
	\expblue{24} & $\emptyset$ & A (20)
	& HHHH (5), {\it AHHH} (3), {\it AAHH} (17), {\it AAAH} (40), AAAA (35) \\
	
	\expblue{25} & $\emptyset$ & A (20), L (20)
	& HHHH (0.5), {\it AHHH} (3), {\it AAHH} (6), {\it AAAH} (3), {\it LHHH} (0.5), {\it AALH} (4), {\it AAAL}+{\it LLHH} (16), {\it ALHH}+AAAA (6), {\it LLAH} (10), {\it LLAA} (23), {\it LLLH} (4), {\it ALLL} (22), LLLL (6) \\
	
	\addlinespace[2em]
\end{longtable}

\section{Theoretical method}
\label{sec:method}

\subsection{Equilibrium and notations}
A reaction in a solution at temperature $T$
\begin{equation}
  \alpha A + \beta B +\dots \rightleftharpoons \rho R + \sigma S+\dots, 
\end{equation}
is characterised by the equilibrium constant
\begin{equation}
  \label{eq:K}
  K=\frac{\lbrack R\rbrack_{\rm eq}^\rho \lbrack S\rbrack_{\rm eq}^\sigma\dots}{\lbrack A\rbrack_{\rm eq}^\alpha \lbrack B\rbrack_{\rm eq}^\beta\dots}= \exp \left(-\frac{\Delta G^\circ}{k_BT}\right),
\end{equation}
%Moreover, $K$ is related to the standard Gibbs energy change $\Delta G^\circ$ of the reaction: 
%\begin{equation}
%  \label{eq:K2}
%  K = \exp \left(-\frac{\Delta G^\circ}{k_BT}\right),
%\end{equation}
with $k_B$ the Boltzmann constant and $\Delta G^\circ$ the standard Gibbs energy change. %(sometimes also misleadingly called the free enthalpy, to be distinguished from the enthalpy $H=G+TS$). 
We are at usual conditions of temperature and pressure, thus the chemical potentials $\mu_a$ of species $a$ in the solution are the standard ones, $\mu_a^0$. 
$\Delta G^\circ$ depends on the chemical potentials: 
\begin{equation}
  \Delta G^\circ = \rho \mu^0_R + \sigma\mu^0_S+\dots - \alpha\mu_A^0 - \beta\mu_B^0 - \dots. 
\end{equation}

\subsection{Application to the experiments}
We now derive the equilibrium associated to a given experiment $\mathcal E = \lbrack \mathcal B, \{c^0_a, a\in\mathcal A\}, \{c_p, p\in\mathcal P^{\rm exp}\} \rbrack$.
First, we compute the list $\mathcal P^{\rm th}$ of peptides present in the medium. 
The number $n_p$ of elements of $\mathcal P^{\rm th}$ depends on the experiment: $n_p=(n_{\rm BB}+1)^{4-|{\mathcal B|}}$ where $|\mathcal B|$ is the cardinal of $\mathcal B$ (for example, $n_p=2$ for the first experiment of Tab.~\ref{tab:experiments}).
Any couple $(p,q)$ of peptides in $\mathcal P^{\rm th}$ is related by a chemical reaction involving amino acids. The equilibrium of the reactive mixture is given by a maximal set of $n_r = n_p - 1$ independent elementary reactions. 
We choose the reactions such that each of them has a single different peptide in the products, different from the backbone, and that the reactants are a peptide and a single amino acid. 

We illustrate these concepts on experiment (13). 
In this case,  $\mathcal B=\{1,8\}$, $\mathcal A=\{A,L\}$, $n_{\rm BB}=2$, $n_p=9$ and the peptides present in the solutions are represented on Fig.~\ref{fig:backbone}(c):
\begin{equation*}
\mathcal P^{\rm th}=\{ {\rm NHHN, NHAN, NAHN, NHLN, NLHN, NAAN, NLLN, NALN, NLAN}\}.
\end{equation*}
If we consider $p=$NLLN and $q=$NHAN, the chemical reaction involving $(p,q)$ is:
\begin{equation}
  \label{eq:reac_peptide}
  \quad {\rm NLLN} + {\rm A} + s \rightleftharpoons {\rm NHAN}  + 2{\rm L}
\end{equation}
where $s$ is a byproduct, which is the same whatever the  scaffold site and whatever the amino acid substituted.~\cite{comment} 
%Note that for readibility, we have oversimplified the reaction: the amino acid in solution is in reality attached to a molecule, represented by a $5$ in Fig.~\ref{fig:spectra}, giving rise to the byproduct when the amino acid is substituted by the sulfur group coming from the backbone, see \cite{chem.202200454} for details.

A possible maximal set of $n_r = 8$ independant elementary reactions is:
%\begin{enumerate}
%  \item $2a + A \rightleftharpoons 2aA^3 + s$, 
%  \item $2a + A \rightleftharpoons 2aA^6 + s$, 
%  \item $2a + 2A \rightleftharpoons 2aA^3A^6 + 2s$, 
%  \item $2a + L \rightleftharpoons 2aL^3 + s$, 
%  \item $2a + L \rightleftharpoons 2aL^6 + s$,
%  \item $2a + 2L \rightleftharpoons 2aL^3L^6 + 2s$,
%  \item $2a + A + L \rightleftharpoons 2aA^3L^6 + 2s$,
%  \item $2a + A + L \rightleftharpoons 2aL^3L^6 + 2s$.
%\end{enumerate} 
\begin{gather}
\label{eq:exemple_K}
  {\rm NHHN + A \rightleftharpoons NAHN} + s, \\
  {\rm NHHN + A \rightleftharpoons NHAN} + s, \\
  {\rm NAHN + A \rightleftharpoons NAAN} + s, \\
  {\rm NHHN + L \rightleftharpoons NLHN} + s, \\
  {\rm NHHN + L \rightleftharpoons NHLN} + s,\\
  {\rm NLHN + L \rightleftharpoons NLLN} + s,\\
  {\rm NAHN + L \rightleftharpoons NALN} + s,\\
  {\rm NLHN + A \rightleftharpoons NLAN} + s.
\end{gather}
 
The standard chemical potentials of the $n_p+n_{\rm BB}+1$ species of the mixture (the elements of $\mathcal P^{\rm th}$, the amino acids in $\mathcal A$ and the byproduct) give us access to the $n_r$ equilibrium constants $K_{i=1\dots n_r}$ and to all the concentrations of the chemical species involved in the experiment, as we now detail. 

The initial backbone concentration (NHHN in experiment (13)) is arbitrarily fixed to $c^0_b=1$, the initial ratio of concentration $c^0_a$ for each amino acid $a\in \mathcal A$ is known, and the byproduct is initially absent: $c_s^0=0$. These $n_{BB}+2$ initial conditions implies $n_{\rm BB}+2$ equations using the matter conservation:
(i) the final backbone concentration is 
$$c_b=c^0_b-\sum_{p\neq b} c_p,$$ 
with $c_p$ the final concentration of peptide $p$,
(ii) the amino acid concentrations $c_a$ are linear combinations of their initial concentrations $c_a^0$ and of the peptides concentrations $c_p$ weighted by the number $n_a(p)$ of $a$ in the peptide, respectively: 
$$c_a=c_a^0 - \sum_p n_a(p)c_p,$$
(iii) the final byproduct concentration $c_s$ is fully determined by the amino acid concentrations:  $$c_s=\sum_{a \in \mathcal A} (c^0_a-c_a).$$

Together with the $n_r = n_p - 1$ equations obtained from Eq.~\eqref{eq:K} for the maximal set of independant reactions (see the exemple of Eq.~\eqref{eq:exemple_K}), these $n_{BB}+2$ constraints form the $n_p+n_{\rm BB}+1$ equations fully characterizing the mixture composition at equilibrium. 

However, the system of equation is non-linear. 
Our choice of elementary reactions, with two reactants and two products: $p_1 + a \rightleftharpoons p_2 +s$, give us a system of $n_r$ quadratic equations in the $c_p$'s:
$K c_{p_1}c_a - c_{p_2}c_s  = 0$, with one such equation per elementary reaction.

With the hypothesis that the chemical potentials are known, the equilibrium constants are also known, and the concentrations of all species. But this is not the case and the number of chemical potentials is large. 
We now derive a model to evaluate the chemical potentials from a small set of parameters.

\subsection{Model for the chemical potential of the peptides}

We propose to express the standard chemical potential of the $n_p=(n_{\rm BB}+1)^{4-|{\mathcal B|}}$ peptides as a function of a reduced number of parameters. 
We represent the amino acids as interacting particles on a 4-site lattice (the scaffold) as illustrated in Fig.~\ref{fig:backbone}(c).
$\mu_p^0$ can be written as a sum of energetic two-body contributions over each couple of sites $(i,j)$ ($i,j=1,3,6,8$), that depend on the groups $(g_i, g_j)$ ($g=$H, N, or an amino acid of $\mathcal A$).
We neglect the triplet and quadruplet coupling terms:
\begin{equation}
  \mu^0_p = \sum_{i\leq j} J^{i,j}_{g_i, g_j}. 
\end{equation}
When $i=j$, the term corresponds to the proper enthalpy of group $g_i$, denoted $E_{g_i}^i$ to lighten notations.  We fix $J^{j,i}_{g_j, g_i}=J^{i,j}_{g_i, g_j}$. 

For example, the energy of the NALN peptide ( bottom right of Fig.~\ref{fig:backbone}(c) writes:
\begin{equation}
\label{eq:mugeneral}
\mu_{\rm NALN}^0=E^1_{N}+E^3_{A}+E^6_{L}+E^8_{N}
+J^{1,3}_{N,A}+J^{1,6}_{N,L}+J^{1,8}_{N,N}+J^{3,6}_{A,L}+J^{3,8}_{A,N}+J^{6,8}_{L,N}
\end{equation}
In each experiment, only the blocked sites are and stay N. Moreover, we can choose the origin of energies at our convenience. Thus we fix for any sites $i\neq j$:
\begin{equation}
\label{eq:ZeroE}
\begin{cases}
E^i_{\rm N}=E^i_{\rm H}=0,\\
E^{i,j}_{\rm N,N}=E^{i,j}_{\rm N,H}=E^{i,j}_{\rm H,H}=0.
\end{cases}
\end{equation}
At this point, for a number $n_{\rm BB}$ of amino acids, our model requires $4n_{\rm BB}+12((n_{\rm BB}+2)^2-3)$ parameters for determining $(n_{\rm BB}+2)^4-2^4$ chemical potentials of peptides of interest. 
In the case of the experiments of Tab.~\ref{tab:experiments}, with two amino acids, it gives 86 parameters against $240$ chemical potentials.

In order to still reduce the number of parameters, we now assume that the reactive face (containing the reactive sites) possesses two mirror symmetry planes, as shown in Fig.~\ref{fig:backbone}(b). 
It implies that: 
\begin{itemize}[itemsep=1pt]
	\item $E_{g}^i$ is independent of $i$. 
  We note it $E_g$.
	\item 
  the two-body couplings reduce to three types: 
  lateral $l$ ($1\leftrightarrow3$, $6\leftrightarrow8$), 
  transverse $t$ ($3\leftrightarrow6$, $1\leftrightarrow8$) and 
  cross $c$ ($1\leftrightarrow 6$, $3\leftrightarrow8$) (represented in Fig.~\ref{fig:backbone}(c)). 
  We change the notation $J^{i,j}_{g_1,g_2}\to L_{g_1,g_2}$ for lateral, $T_{g_1,g_2}$ for transverse and $C_{g_1,g_2}$ for cross links.
\end{itemize}
The choice of reference energy in Eq.~\eqref{eq:ZeroE} implies in the new notations:
\begin{equation}
\begin{cases}
E_{\rm N}=E_{\rm H}=0,\\
X_{\rm N,N}=X_{\rm H,H}=X_{\rm N,H}=0, \qquad X=L,T,C
\end{cases}
\end{equation}  
Finally, assuming that the coupling is dominated by steric repulsion and considering that H is  much smaller that any amino acid, we identify the coupling between an amino acid denoted $x$ and a non reactive site H with the coupling between (H,H):
\begin{equation}
T_{x,H}=L_{x,H}=C_{x,H}=0. 
\end{equation}
The energy of the 'nude' scaffold is thus equal to zero and taken as a reference.
The chemical potential of Eq.~(\ref{eq:mugeneral}) is expressed as follows, given the aforementioned assumptions:
\begin{equation}
\label{musym}
\mu_{\rm NALN}^0=E_{\rm A}+E_{\rm L}+T_{\rm A,L}+L_{\rm A,N}+L_{\rm L,N}+C_{\rm L,N}+C_{\rm A,N}
\end{equation}
The parameters involved in each experiment of Tab.~\ref{tab:experiments} are summarized in Tab.~\ref{tab:parametersPresence}. 

\begin{table}
\begin{tabular}{c|c|c|c|c|c|c|c|c|c|c|c|c|c|c|c|c|c}
Exp& $E_A$ & $E_L$ & $T_{AA}$ &  $T_{AL}$ &  $T_{AN}$ &  $T_{LL}$ &  $T_{LN}$ &  $L_{AA}$ &  $L_{AL}$ & $L_{AN}$ &  $L_{LL}$ &  $L_{LN}$ & $C_{AA}$ &  $C_{AL}$ &  $C_{AN}$ &  $C_{LL}$ &  $C_{LN}$
\\ \hline
(1-3) &-&\chk&-&-&-&-& \chk&-&-&-&-& \chk&-&-&-&-&\chk\\
\hline
(4-6) &\chk&-&-&-&\chk&-&-& -&-&\chk&-&-& -&-&\chk&-&- \\ 
\hline \hline
(7-9) &-&\chk&-&-&-&\chk& -&-&-&-&-& \chk&-&-&-&-&\chk\\
%(4) &-&\chk&-&-&-&\chk& -&-&-&-&-& \chk&-&-&-&-&\chk\\
%(5) &-&\chk&-&-&-&\chk& -&-&-&-&-& \chk&-&-&-&-&\chk\\
\hline
(10-11) &\chk&-&\chk&-&-&-&-& -&-&\chk&-&-& -&-&\chk&-&- \\ 
\hline
%(7) &\chk&-&\chk&-&-&-&-& -&-&\chk&-&-& -&-&\chk&-&-\\
%(8) &\chk&-&\chk&-&-&-&-& -&-&\chk&-&-& -&-&\chk&-&-\\ \hline
(12-14) &\chk&\chk&\chk&\chk&-&\chk&-& -&-&\chk&-&\chk&-&-&\chk&-&\chk\\
\hline \hline
(15-17) &-&\chk&-&-&-&-&\chk&  -&-&-&-& \chk&-&-&- &\chk&-\\
\hline
(18-20) &\chk&-&-&-&\chk&-&-& -&-&\chk&-& 
- &\textcolor{red}{\checkmark} & -&-&-&- \\ 
%(13) &\chk&-&-&-&\chk&-&-& -&-&\chk&-& 
%-&\textcolor{red}{\checkmark} & -&-&-&- \\ 
%(14) &\chk&-&-&-&\chk&-&-& -&-&\chk&-& 
%-&\textcolor{red}{\checkmark} & -&-&-&- \\ 
%\hline 
%
%(16) &-&\chk&-&-&-&-&\chk&  -&-&-&-& \chk&-&-&- &\chk& -\\
%(17) &-&\chk&-&-&-&-&\chk&  -&-&-&-& \chk&-&-&- &\chk& -\\
\hline
(21) &\chk&\chk&-&-&\chk&-&\chk&  -&-&\chk&-& \chk&\chk&\chk&- &\chk& -\\
%(19) &\chk&\chk&-&-&\chk&-&\chk&  -&-&\chk&-& \chk&\chk&\chk&- &\chk& -\\
%(20) &\chk&\chk&-&-&\chk&-&\chk&  -&-&\chk&-& \chk&\chk&\chk&- &\chk& -\\
\hline \hline
(22-23) &-&\chk&-&-&- & \chk&-&-&-&- & \chk&-&-&-&- &\chk&-\\
\hline
(24) &\chk&-& \chk&-&-&-&-&\chk&-&- &-& 
- &\textcolor{red}{\checkmark} & -&-&-&- \\ 
\hline
%(26-27) &-&\chk&-&-&- & \chk&-&-&-&- & \chk&-&-&-&- &\chk&-\\
%(23) &-&\chk&-&-&- & \chk&-&-&-&- & \chk&-&-&-&- &\chk&-\\

(25) &\chk&\chk&\chk&\chk&- & \chk&-& \textcolor{red}{\checkmark}&\chk&- & \chk&-&\chk&\chk&- &\chk&-\\
\end{tabular}
\caption{List of the interactions present in each experiment of Tab.~\ref{tab:experiments}. 
Double lines separate experiments with different scaffolds. }
\label{tab:parametersPresence}
\end{table}

These assumptions reduce the number of parameters that must be inferred to treat all the experiments in Tab.~\ref{tab:experiments} (i.e. for any scaffold and 2 amino acids). Without making any assumptions about the symmetry of the system, we have $2\times 4+(4\times3)/2\times (4\times 4-3)=86$ parameters, where $4\times 2$ refers to the 4 proper energies for the 2 amino acids, $(4\times3)/2$ corresponds to all the site pairs and $(4\times 4-3)$ all the BB pairs in (A,L,N,H), when subtracting HH, NH, NN for which pair interactions are set to 0. After simplification of the system, we get  $2+3\times 5=17$: 2 proper energies ($E_A$ and $E_L$), 5 lateral, cross, and transverse terms ($X_{AA},X_{AL},X_{LL}, X_{NA}, X_{NL}$). 
%\lmc{
%Les énergies propres, ça fait 2*4, et pas 4*4, sauf si on ne choisit pas la référence d'énergie, mais ça, c'est pas une approx. 
%Je dirais 2*4+(4*3)/2*(4*4-3)=86. Le (4*3)/2 c'est pour tous les couples de sites différents, et 4*4-3 c'est pour tous les doublets avec éléments parmi (A,L,N,H), mais on enlève HH, NH, NN.}

More generally, for $n_a$ different amino acids, the number of parameters after simplification is $n_a+3\left(\frac{(n_a+2)(n_a+1)}2-1\right)$, reaching 710 parameters for $n_a=20$. 
The objective is now to design a procedure to determine the parameters from a reduced set of experiments, allowing to determine the concentrations of other experiments. 

\subsection{Inference of the parameters}

The number of chemical potentials required to determine the concentrations in any experiment involving a given set of $n_{\rm B}$ amino acids is the number of possible species: $n_p + n_{BB} + 1$. 
But $\mu^0_s$ for the byproduct of Eq.~\eqref{eq:reac_peptide} and $\mu_a^0, a\in \mathcal A$ can be set to zero by adjusting the chemical potentials of the peptides. 
In the model of the previous section, it is possible simply adjusting $E_A$. 
Thus the parameters of our model are enough to calculate all the concentrations of an experiment set. 

We denote $\mathcal X$ a set of parameters: $\mathcal X=\{E_g; g\in \mathcal A\}\cup \{X_{g_i,g_j}; X=L,T,C; (g_i,g_j\neq g_i) \in (\mathcal A\cup\{N\})^2/\{(N,N)\}\}$. 

In the case of Tab.~\ref{tab:experiments}, the set of experiments furnishes 102 data, corresponding to the relative intensities of 102 peaks from HPLC experiments, to compare with 17 model parameters.

%\subsection{Main steps}
We choose any subset of the $n_{\rm exp}=25$ experiments of Tab.~\ref{tab:experiments} involving one or two amino acids among $\{A,L\}$.
$\mathcal P_i^{\rm th}$ is the set of peptides that can be synthesized in experiment $i$ and $\mathcal P=\cup_{i=1}^{n_{\rm exp}} \mathcal P^{\rm th}_i$ is their union. 
Neglecting the approximations of our model and the experimental errors, the correct $\mathcal X$  gives exactly the experimental measurements. 
Thus, our objective will be to minimize some distance between the concentrations $c^{\mathcal X}_p$ obtained from $\mathcal X$ for the set of experiments and those $c^{\rm exp}_p$ of the set of experiments, departing from random values for $\mathcal X$. 
We define a distance $\Delta_{\mathcal X}$ that measures the distance between the theoretical predictions and the experimentally measured values of the mixture compositions. We choose it as the sum over all the peaks obtained in the experimental subset of the squared logarithm of the ratio between the peak area obtained from $\mathcal X$ and from the experiments.

The steps are 
\begin{enumerate}
  \item Determine the $\mathcal P_i$'s from the set of experiments:
  $$\mathcal E_i\to \mathcal P_i.$$
  
%  model.allPeptidesNb( exp[0], exp[1])
  \item Determine the chemical potentials for each peptide of $\mathcal P_i$ and a trial set of parameters:
  $$(\mathcal X, \mathcal P_i) \to \{\mu_p^{\mathcal X}, p\in \mathcal P_i\}.$$
%  model.CalcMuP(allPepNb, vInteraction0)
  \item Determine the theoretical concentrations for each experiment from the chemical potentials:
  $$(\mathcal P_i, \{\mu_p^{\mathcal X}, p\in \mathcal P_i\}, \mathcal E_i)\to \{c^{\mathcal X,i}_{p}, p\in \mathcal P_i\}.$$
  \item Determine a distance $\Delta_{\mathcal X}$ between the concentrations obtained from $\mathcal X$ and from the experiments:
  $$\left\{\lbrack \mathcal E_i, \{c^{\mathcal E_i}_p-c^{\mathcal X}_p , p\in \mathcal P_i\}\rbrack, i=1..n_{\rm exp}\right\}
  \to \Delta_{\mathcal X}$$
\end{enumerate}
  We introduce a distance function defined as:
  \begin{equation}
    \Delta_{\mathcal X} = \sum^{n_{\rm exp}}_{l=1}\sum_{{\rm peaks } \,k} \left(\log(\frac{ c^{\mathcal X,i}_{k}}{ c^{\mathcal E_l}_k})\right)^2
  \end{equation}
where $c^{\mathcal E_l}_k$ is the relative area of the $k^{th}$ peak of HPLC analysis of the $l^{th}$ experiment. It is equal to the proportion of one species $j$ or of a mix of species that the analysis method can not resolve. $c^{\mathcal X,i}_k$ is the sum of the theoretically computed proportions $\Sigma_{k=1}^K  c^i_{{\rm theo}, k}$  of the $K$ species detected in the $j^{th}$ peak. Note that we choose a squared logarithmic distance because peak concentrations span several orders of magnitude, so the metric weights relative errors symmetrically — a prediction off by a factor of 2 is penalized the same as one off by a factor of 1/2 — which is the relevant notion of error for compositional data.

% $\Delta a$ is the standard deviation of the experimental measure $a^i_{{\rm exp}, j}$. 

%\lms{The distance function, $ \Delta_{\mathcal X}$, measures the distance between the theoretical predictions and the experimentally measured values of the mixture compositions.   The smaller the distance, the better the model reproduces the observations.}

We minimize $ \Delta_{\mathcal X}$ using two methods:
\begin{enumerate}
  \item a random walk minimization which step is labeled as ($t=1,...,1000$). At step $t$, we draw a set of 17 random values for $\mathcal X_t$, compute $ \Delta_{\mathcal X}^t$, if  $\Delta_{\mathcal X}^t< \Delta_{\mathcal X}^{t-1}$ we replace the set  $\mathcal X_{t-1}$ by $\mathcal X_t$, otherwise we keep  $\mathcal X_{t-1}$  and iterate.
  \item a function minimize of numpy python package using the "Powell" descent gradients methods. It returns the set $\mathcal X$ giving the minimum values for $\Delta_{\mathcal X}$.
\end{enumerate}

The two methods give comparable results. 
We focus in the second one to compute the results. We perform the minimisation ten times from ten different starting parameter sets $\mathcal X_{0}$, randomly generated.

\section{Results and discussion}
\subsection{Validation of the model}
\label{sec:validation}
To validate our approach, we split the experiments presented in Tab.~\ref{tab:experiments} into a "training group"-18 experiments, so three-fourths of the set, labelled in blue in Tab.~\ref{tab:experiments} -and a validation group -7 experiments, so about one fourth, labelled in red. The training experiments are used to infer the parameters $\mathcal X$, that we divide into subsets of parameters $E$ (proper energies), $L$ (lateral coupling), $T$ (transverse coupling) and $C$ (cross coupling). We use the obtained parameters to predict the composition of the validation group and compare the prediction to experimental data.

To evaluate the accuracy of the model, we progressively enlarge our set of optimized parameters: first, we only consider the proper energies $E_i$ ($\mathcal X=E$) and compute and mimize the  distance $\Delta \chi$ for the training experiments following the steps described in the previous section.
Then, we repeat the optimization when including the transverse coupling terms, $T_{ij}$, occurring between sites (1, 8), and (3, 6), as shown in Fig.~\ref{fig:backbone}(c): $\mathcal X=(E,T)$. These are the closest neighbours in the backbone, so we assume that they are the dominant coupling terms. 
Next, we  add lateral coupling parameters between sites (1, 3) and (6, 8): $\mathcal X=(E,T,L)$. 
and finally, we include the coupling between cross-sites (1, 6) and (3, 8), which correspond to the largest distance between amino acids in the backbone. 
The optimized parameter values for each case are listed in Tab.~\ref{validation_tab}. We perform the optimization for 10 initial sets $\mathcal X_0$ 
with values randomly distributed between -5 and 5. The minimisation converges to the same points, independently of the initial values, regardless of negligible variations (as illustrated in Appendix A).
\begin{table}[ht]
	\centering
	\footnotesize
	\setlength{\tabcolsep}{4pt}
	\begin{tabular}{c c c c c c c c}
		\toprule
		$\mathcal X$ & $i$ & $E_j$ & $i,j$ & $T_{i,j}$ & $L_{i,j}$ & $C_{i,j}$ & $\Delta_\chi$ \\
		\midrule
		
		\textbf{$E$} &
		\begin{tabular}{c}
			$A$\\
			$L$
		\end{tabular} &
		\begin{tabular}{c}
			1.022\\
			0.916
		\end{tabular} &
		\begin{tabular}{c}
			$NA$\\
			$NL$\\
			$AA$\\
			$AL$\\
			$LL$
		\end{tabular} &
		\begin{tabular}{c}
			0\\0\\0\\0\\0
		\end{tabular} &
		\begin{tabular}{c}
			0\\0\\0\\0\\0
		\end{tabular} &
		\begin{tabular}{c}
			0\\0\\0\\0\\0
		\end{tabular} &
		72.8 \\
		\cmidrule(lr){4-8}
		
		\textbf{$(E,T)$} &
		\begin{tabular}{c}
			$A$\\
			$L$
		\end{tabular} &
		\begin{tabular}{c}
			2.182\\
			2.041
		\end{tabular} &
		\begin{tabular}{c}
			$NA$\\
			$NL$\\
			$AA$\\
			$AL$\\
			$LL$
		\end{tabular} &
		\begin{tabular}{c}
			-2.339\\
			-2.487\\
			-1.441\\
			-1.257\\
			-1.742
		\end{tabular} &
		\begin{tabular}{c}
			0\\0\\0\\0\\0
		\end{tabular} &
		\begin{tabular}{c}
			0\\0\\0\\0\\0
		\end{tabular} &
		40.1 \\
		\cmidrule(lr){4-8}
		
		\textbf{$(E,T,L)$} &
		\begin{tabular}{c}
			$A$\\
			$L$
		\end{tabular} &
		\begin{tabular}{c}
			2.502\\
			3.209
		\end{tabular} &
		\begin{tabular}{c}
			$NA$\\
			$NL$\\
			$AA$\\
			$AL$\\
			$LL$
		\end{tabular} &
		\begin{tabular}{c}
			-0.393\\
			-1.030\\
			-0.094\\
			-0.414\\
			-1.214
		\end{tabular} &
		\begin{tabular}{c}
			-2.191\\
			-2.704\\
			-1.6966\\
			-1.598\\
			-2.083
		\end{tabular} &
		\begin{tabular}{c}
			0\\0\\0\\0\\0
		\end{tabular} &
		7.6 \\
		\cmidrule(lr){4-8}
		
		\textbf{$(E,T,L,C)$} &
		\begin{tabular}{c}
			$A$\\
			$L$
		\end{tabular} &
		\begin{tabular}{c}
			2.523\\
			3.211
		\end{tabular} &
		\begin{tabular}{c}
			$NA$\\
			$NL$\\
			$AA$\\
			$AL$\\
			$LL$
		\end{tabular} &
		\begin{tabular}{c}
			-0.318\\
			-0.682\\
			-0.138\\
			-0.309\\
			-1.046
		\end{tabular} &
		\begin{tabular}{c}
			-2.331\\
			-2.548\\
			-1.801\\
			-1.631\\
			-1.133
		\end{tabular} &
		\begin{tabular}{c}
			0.130\\
			-0.252\\
			0.112\\
			-0.092\\
			-1.095
		\end{tabular} &
		6.7 \\
		
		\bottomrule
	\end{tabular}

\vspace{0.8em}
\caption{Values for dimensionless parameters $E_i$, $T_{ij}$, $L_{ij}$, $C_{ij}$, optimising the reproduction of the training expriments, indicated in blue in Tab.~\ref{tab:experiments}. We increase the complexity of the model for the peptide chemical potential from pure proper energy to the full set of coupling parameters.
}
\label{validation_tab}
\end{table}

The distance function, $\Delta_\chi$, in the last column of Tab.~\ref{validation_tab} decreases as the model becomes more complex. It can be seen that considering transverse and lateral coupling significantly decreases the distance $\Delta_{\mathcal X}$, whereas introducing the cross terms has a minor effect. We also notice that the proper energies are positive, whereas the coupling terms are are negative or vanishing.

\begin{figure}
\begin{center}
\includegraphics[width=0.7\textwidth]{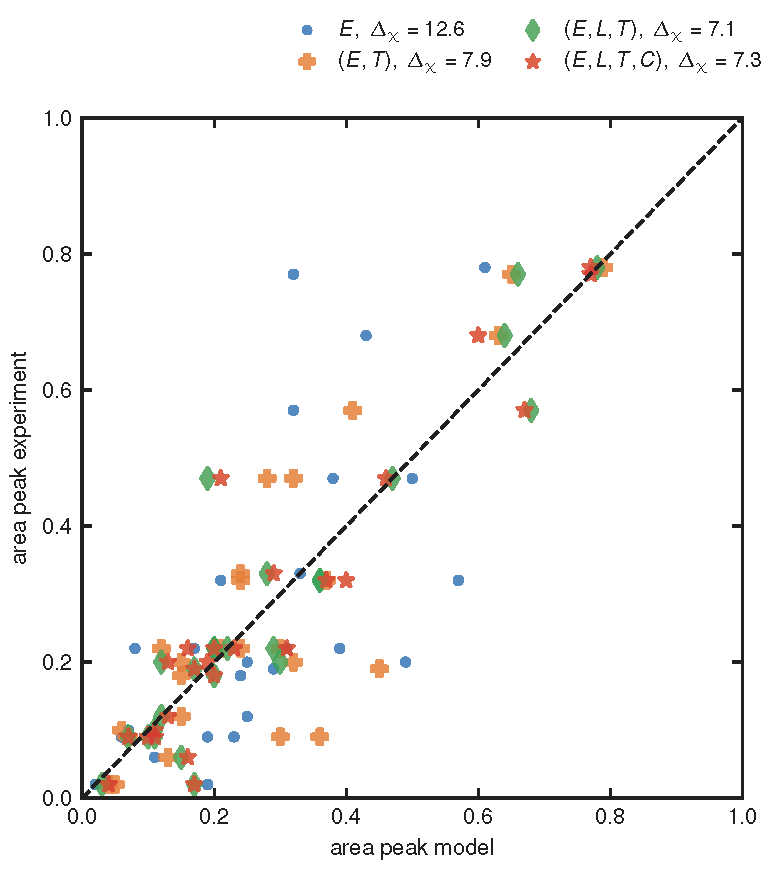}
\caption{ Experimental peak area for experiments validation (red in Tab.~\ref{tab:experiments}) as a function of the one obtained with the model. Blue circles correspond to model with proper energy $E$, orange cross to the $(E,T)$ model, green plus to the $(E,T,L)$ model and red stars to $(E,T,L,C)$ model. The dashed line corresponds to the $y=x$ line indicating perfect agreement between experiments and prediction.}
\label{fig:correlation}
\end{center}
\end{figure}

Now, we use the optimised set of parameters (optimised for the training experiments) to calculate the composition of the validation experiments. As before, we increase the complexity of the model step by step (first only proper energy, then proper energy plus transverse coupling, and so on) and calculate the relative area of the peaks in the validation experiments for each model. 
The results, displayed in Fig.~\ref{fig:correlation}, show the experimental peak area as a function of the modeled area for each optimisation parameter-set. 
All the points form a cloud roughly along the $y = x$ line, thus validating the global approach of this work. 
More precisely, the predictions improve when moving from $\mathcal X=E$ (blue circles) to proper energy and coupling terms. However, switching from $\mathcal X=(E,T,L)$ to $\mathcal X=(E,T,L,C)$ only marginally improves the prediction, as can be seen in the plot where most of the green triangles corresponding to $\mathcal X=(E,T,L)$ and the red stars superpose. This is confirmed by the value of $\Delta_{\mathcal X}$: $\Delta_{(E)}=12.6$, $\Delta_{(E,T)}=7.9$, $\Delta_{(E,T,L)}=7.1$ and finally, $\Delta_{(E,T,L,C)}=7.3$.  Note that the distance remains non-vanishing, partly because data are not fully consistent. In particular, the experiment (23) reports a maximum probability for an incomplete reaction for the leucine (LLLH, more concentrated than LLLL) in contradiction with most of the data. This leads to predictions that remain far from experiments and points offset from the $x=y$ line. 
To quantify the typical error of each model, we computed the median factor, defined as $\exp( {\rm median} |\log(y_i/x_i)|)$, where $y_i$ and $x_i$ are the predicted and experimental area of peak $i$. The median factor describes the error of a typical data point : half of the predictions deviate from experiment by less than this factor. Here, the median factor drops from 1.53 for  $\mathcal X=E$  to 1.13 for $\mathcal X=(E,T,L)$  and $\mathcal X=(E,T,L,C)$. It means that half of the peaks are predicted within ~13\% relative error. This shows that the method gives a very good prediction for most of the data even though the distance remains elevated due to a few outliers.

\subsection{Results for the all data}
Now that we have validated our model, the final section is devoted to determining the composition of the DCL.
We apply the inference procedure to the entire set of experiments of Tab.~\ref{tab:experiments}, allowing all the interactions.
After a minimization procedure we obtained the set of parameters given in Tab.~\ref{tab:fullresults}.

\begin{table}[h!]
\centering
\begin{tabular}{c|c}
	$i$ & $E_i$\\
	\hline
	$A$ & 2.274\\
	$L$ & 2.757\\
\end{tabular}
\qquad
\begin{tabular}{c|ccc}
	$i,j$ & $T_{i,j}$ & $L_{i,j}$ & $C_{i,j}$ \\
	\hline
	$NA$ & -0.316 & -2.144 & 0.203\\
	$NL$ & -0.690 & -2.288 & -0.129\\
	$AA$ & -0.107 & -1.494 & 0.172\\
	$AL$ & -0.227 & -1.297 & -0.061\\
	$LL$ & -0.881 & -0.884 & -0.871\\
\end{tabular}
\quad
\begin{tabular}{c}
$\Delta_\chi$\\
	\hline
12.94
\end{tabular}
\qquad
\caption{$(E,T,L,C)$ optimizing the full set of experiments (1-25).}
\label{tab:fullresults}
\end{table}
First, we use this set of values to determine the deconvoluted concentrations of the peptides for the experiments (1-9). The results, presented in Appendix B, are convincing. 
Thus, we apply the protocol to determine the concentrations associated with the experiment 25, which is the most complex of our set. Results are given in Tab.~\ref{tab:molecule_concentrations}.

By analysing the table of energy, Tab.~\ref{tab:fullresults}, we can see that the two proper-energies are of the same order of magnitude. They are positive whereas all of the coupling parameters, except for one, are negative. 
This means that couplings stabilize the energy of the peptide formed after the amino acid reacts. The lateral (L) couplings are significantly larger in absolute value than the transverse (T) couplings, reflecting less steric repulsion for lateral interactions. This difference is more pronounced for alanine, a small amino acid that interacts minimally with lateral groups and experiences repulsion primarily with terminal groups, than for leucine, which is bulkier and exhibits similar repulsion across all configurations. Cross-coupling terms are relatively small, serving mainly for fine-tuning and could potentially be neglected.
Each amino acid attached to the backbone involves one proper-energy and three coupling energies. The table~\ref{tab:fullresults} shows that the sum of these parameters is approximately zero, and leucine and alanine behave similarly. Consequently, species containing only alanine or only leucine should have similar concentrations. However, while leucine shows spatial indifference, alanine favors lateral coupling.

	\begin{table}[htbp]
		\centering
		\caption{Concentrations for all peptides synthesised in experiment 25. 
			The first column labels the molecules of equal concentrations for symmetry reasons, the second column presents the concentration of individual species and the last column, the total concentration of the group.
			Purple lines correspond to compounds with alanine, green lines to compounds with leucine and white lines to mixed ones. }
		\label{tab:molecule_concentrations}

	\begin{tabular}{l c c}
		\hline\hline
		\textbf{Molecules} & \textbf{Concentration of one species} & \textbf{total  Concentration} \\
		\hline
		
		\Lgroup
		LLLL (n=1) & 7.97\% & 7.97\% \\
		
		\Nocolor
		LLAA, AALL (n=2) & 1.62\% & 3.24\% \\
		
		\Nocolor
		LLLA, LLAL, LALL, ALLL (n=4) & 4.00\% & 16.00\% \\
		
		\Agroup
		AAAA (n=1) & 4.94\% & 4.94\% \\
		
		\Nocolor
		LAAA, ALAA, AALA, AAAL (n=4) & 3.14\% & 12.56\% \\
		
		\Nocolor
		LAAL, ALLA (n=2) & 3.42\% & 6.84\% \\
		
		\Lgroup
		LLLH, LLHL, LHLL, HLLL (n=4) & 0.94\% & 3.76\% \\
		
		\Nocolor
		LALA, ALAL (n=2) & 4.55\% & 9.10\% \\
		
		\Nocolor
		LLHA, LLAH, HALL, AHLL (n=4) & 0.31\% & 1.24\% \\
		
		\Lgroup
		LHHL, HLLH (n=2) & 0.27\% & 0.54\% \\
		
		\Nocolor
		LHAA, HLAA, AALH, AAHL (n=4) & 0.93\% & 3.72\% \\
		
		\Nocolor
		LHAL, LAHL, HLLA, ALLH (n=4) & 0.91\% & 3.64\% \\
		
		\Agroup
		HAAA, AHAA, AAHA, AAAH (n=4) & 1.20\% & 4.80\% \\
		
		\Nocolor
		LHLA, LALH, HLAL, ALHL (n=4) & 1.20\% & 4.80\% \\
		
		\Nocolor
		LHHA, HLAH, HALH, AHHL (n=4) & 0.23\% & 0.92\% \\
		
		\Nocolor
		LAAH, HAAL, ALHA, AHLA (n=4) & 0.68\% & 2.72\% \\
		
		\Lgroup
		LLHH, HHLL (n=2) & 0.27\% & 0.54\% \\
		
		\Nocolor
		LAHA, HALA, ALAH, AHAL (n=4) & 0.69\% & 2.76\% \\
		
		\Agroup
		HAAH, AHHA (n=2) & 0.33\% & 0.66\% \\
		
		\Agroup
		HHAA, AAHH (n=2) & 1.31\% & 2.62\% \\
		
		\Nocolor
		HHHH (n=1) & 0.31\% & 0.31\% \\
		
		\Lgroup
		LHHH, HLHH, HHLH, HHHL (n=4) & 0.18\% & 0.72\% \\
		
		\Lgroup
		LHLH, HLHL (n=2) & 0.27\% & 0.54\% \\
		
		\Nocolor
		LAHH, HHLA, HHAL, ALHH (n=4) & 0.66\% & 2.64\% \\
		
		\Agroup
		HHHA, HHAH, HAHH, AHHH (n=4) & 0.30\% & 1.20\% \\
		
		\Nocolor
		LHAH, HLHA, HAHL, AHLH (n=4) & 0.17\% & 0.68\% \\
		
		\Agroup
		HAHA, AHAH (n=2) & 0.25\% & 0.50\% \\
		
		\hline\hline
	\end{tabular}

	\end{table}
	\subsection{Discussion}	
Here, we have presented an efficient framework to reconstruct the composition of a complex combinatorial mixture from partial experimental data. We demonstrate its capabilities through the complete interpretation of the chromatogram presented Fig.~\ref{fig:spectra} which is given in Tab.~\ref{tab:molecule_concentrations}.

A key strength of our approach is the dramatic reduction in the number of parameters required to describe the system. For a mixture with 20 building blocks bearing the side-chains of all natural amino acids, the system contains around 200 000 possible species, yet our approach requires only 710 parameters, a reduction by a factor of approximately 300 compared with the number of possible molecules.  Moreover, the number of parameters introduced in the model can be tuned according to the desired level of detail. For example, we can include site-dependent self-energies that will increase it, while a uniform coupling constant for all geometries (e.g., $T_{XY}$ = $L_{XY}$  = $C_{XY}$) decreases it. This versatility makes the method particularly well-suited for tackling complex mixtures for which only partial information is available, focusing on chemical interactions leading the assembly.

The multivalent architecture of the system studied here provides a particularly demanding test of this approach. With four reactive sites on a common backbone, the possible arrangements of building blocks generate a large combinatorial space, in which the energetic contributions depend not only on the identity of the building blocks but also on their positions and interactions with neighboring units. Our model makes it possible to disentangle these contributions and quantify the role of pairwise couplings in determining the equilibrium composition. In particular, the results reveal that interactions between neighboring building blocks play a major role in the assembly energetics. More generally, this illustrates how a compact physical description can capture the effects of multivalent interactions without requiring an independent parameter for every possible compound.

Another advantage of this deconvolution method is that it simplifies the design and characterization of a library. The full reactivity of the self-assembled mixture (affinity of each BB with the backbone) can be extracted from very complex experiments (in the case presented here, Exp. 25 in Tab.~\ref{tab:experiments} with the full deprotected backbone HHHH, reacting with A and L). If more data are needed for sufficient deconvolution, they can simply be performed at different amino acid concentrations. This greatly simplifies library preparation and enables more practical self-assembly protocols.

\section{Conclusion}

Label discrimination is a common issue in self-assembly processes, which can involve hundreds of species. This limitation hinders the use of self-assembled libraries for identifying new active compounds. Here, we develop a new theoretical method to determine the composition of such systems from partial characterization. Since each compound is an assembly of a small number of building blocks, we express the free energy of each compound in the mixture using a reduced number of parameters. We couple this model with the law of mass action to obtain the concentration of each species. We then optimize the parameter set to maximize the likelihood with the available partial data. As an illustration of the method’s efficiency, we fully resolve the composition of a DCL with 81 species from only 13 HPLC peaks—a task impossible for linear methods.

\section*{Acknowledgments}
HB thanks Olivier Rivoire and Ludovic Jullien for fruitful discussions. We thank the Mass Spectrometry MS3U and the SPPS Platforms (Fédération de chimie moléculaire de Paris centre, FR2769, Sorbonne Université) for services. We acknowledge the doctoral program Interfaces Pour le Vivant for the PhD funding of TP.

\appendix
\section{Convergence of the minimisation procedure }
We illustrate the convergence of the minimisation prossess. Here, we consider the set of test experiments (in blue in Tab.~\ref{tab:experiments}), and we consider only transverse coupling.  We consider 10 different starting sets of parameters $\mathcal X_0$, for which $E_i$ and $T_{ij}$ are randomly distributed between -5 and 5 and we determine $\mathcal X$ minimizing the distance $\Delta_{\mathcal X}$.  The figure represents the dispersion of the final parameters with respect their mean. It is always below 1\%. 
\begin{figure}
	\begin{center}
\includegraphics[scale=0.5]{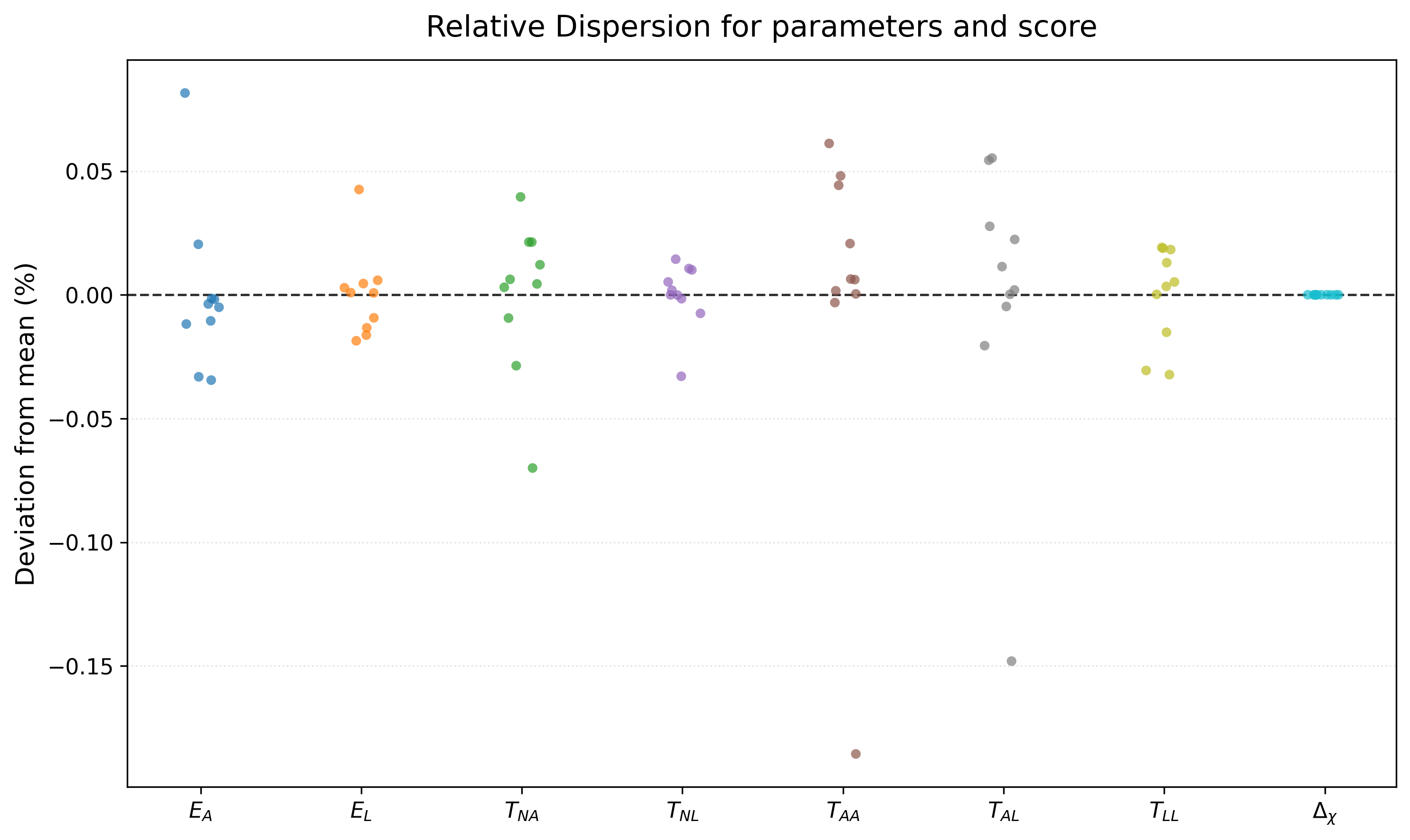}
\caption{Dispersion of the parameters and distance obtained from 10 different starting set $\mathcal{X}_0$. The data presented are obtained for the test experiments (in blue in Tab.~\ref{tab:fullresults})}

\end{center}
\end{figure}

\section{Validation for the full data model }
We plot the comparison between experimental measurements (column $E_i$) and predictions (column $M_i$) for the experiments ($i$=1-9) with the parameters given in Tab.~\ref{tab:fullresults}. Each color block corresponds to a peptide. As one sees the model gives access to the experimental concentration with a precision of a numerical factor of 1.5 in the worst case (9th column, orange and light green).
\begin{figure}
\begin{center}
\includegraphics[scale=0.5]{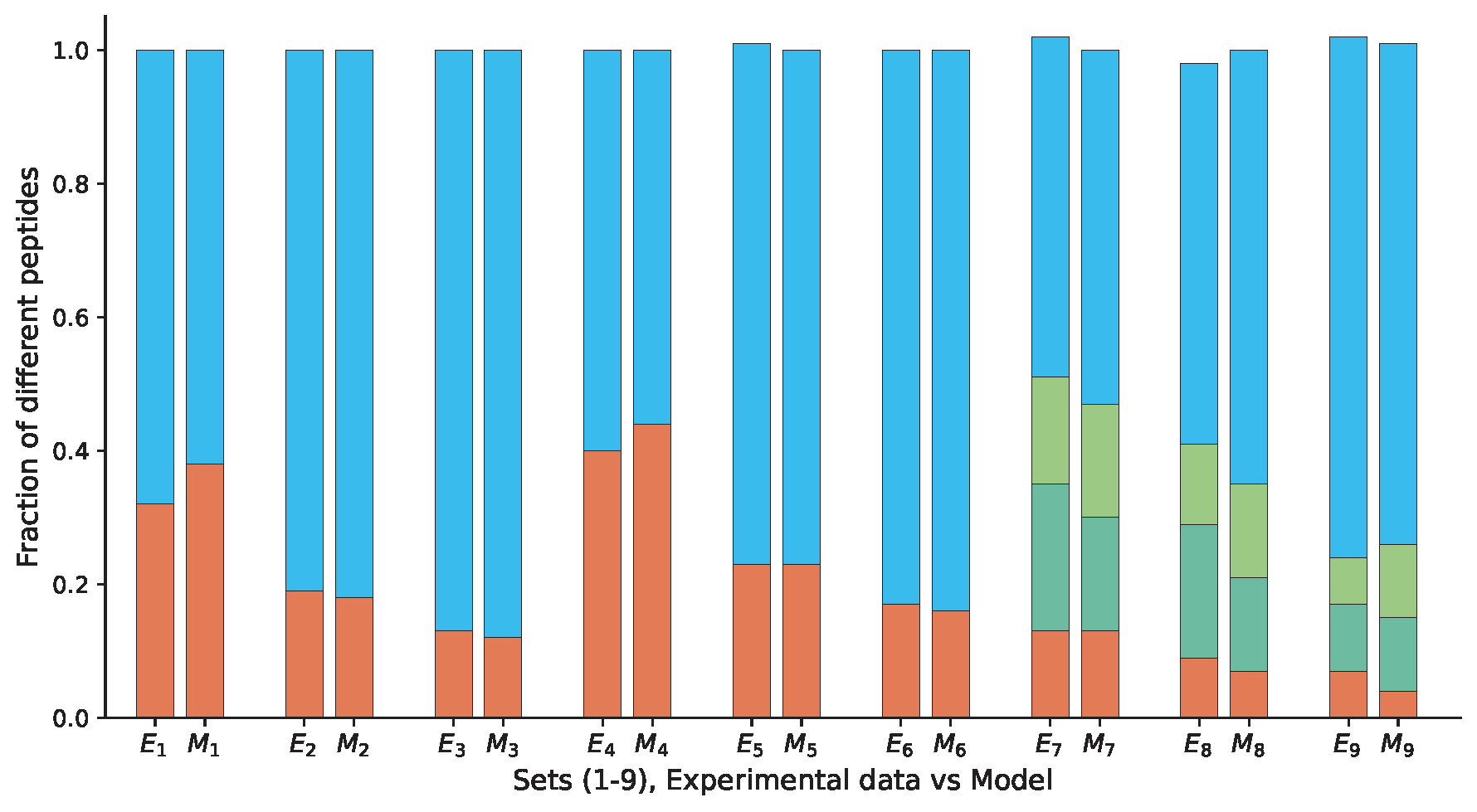}
\caption{Comparison between experimental measurements (column $E_i$) and predictions (column $M_i$) for the experiments ($i$=1-9) with the parameters given in Tab.~\ref{tab:fullresults}. Each color block corresponds to a peptide. }

\label{fig:simple_exp}
\end{center}
\end{figure}

\bibliography{peptides.bib}

\begin{thebibliography}{25}
\bibitem [{com()}]{comment}%
  \BibitemOpen
  \href@noop {} {\bibinfo {title} {Note that for readibility, we have
  oversimplified the reaction: the amino acid in solution is in reality
  attached to a molecule, represented by a $5$ in fig.~\ref{fig:spectra},
  giving rise to the byproduct when the amino acid is substituted by the sulfur
  group coming from the backbone, see \cite{chem.202200454} for
  details.}}\BibitemShut {Stop}%
\end{thebibliography}
\end{document}